\documentclass{IEEEtran}
\usepackage[hyphens]{url}
\usepackage{graphicx}
\usepackage{xcolor}
\usepackage{colortbl} 
\usepackage{caption}
\usepackage{subcaption}
\usepackage[bookmarks=false]{hyperref}
\usepackage{algorithm}
\usepackage{algorithmicx} 
\usepackage{algpseudocode}
\algnewcommand\algorithmicforeach{\textbf{for each}}
\algdef{S}[FOR]{ForEach}[1]{\algorithmicforeach\ #1\ \algorithmicdo}
\usepackage{hyperref}
\hypersetup{colorlinks=true,linkcolor=black,citecolor=blue,filecolor=black,urlcolor=blue}

\definecolor{headcolor}{gray}{0.9}

\algblock{Input}{EndInput}
\algnotext{EndInput}
\algblock{Output}{EndOutput}
\algnotext{EndOutput}
\newcommand{\Desc}[2]{\State \makebox[2em][l]{#1}#2}
\begin{document}

\title{Performance Evaluation of Big Data Processing Strategies for Neuroimaging}
\pagenumbering{gobble}

\author{
  \IEEEauthorblockN{
    Val\'erie Hayot-Sasson$^1$, Shawn T Brown$^2$ and 
    Tristan Glatard$^1$
  }\\
  \IEEEauthorblockA{
    $^1$Department of Computer Science and Software Engineering, Concordia University, Montreal, Canada\\
    $^2$Montreal Neurological Institute, McGill University, Montreal, Canada
  }
}

\maketitle

\begin{abstract}
    Neuroimaging datasets are rapidly growing in size as a result of 
    advancements in image acquisition methods, open-science and data sharing. 
    However, the adoption of Big Data processing strategies by neuroimaging
    processing engines remains limited. Here, we 
    evaluate three Big Data processing strategies (in-memory computing, 
    data locality and lazy evaluation) on typical neuroimaging use 
    cases, represented by the BigBrain dataset. We contrast these various 
    strategies using Apache Spark and Nipype as our 
    representative Big Data and neuroimaging processing engines, on Dell EMC's 
    Top-500 cluster. 
    Big Data thresholds were modeled by comparing the data-write rate of the 
    application to the filesystem bandwidth and number of concurrent processes. 
    This model acknowledges the 
fact that page caching provided by the Linux kernel is critical to the 
performance of Big Data applications. Results show that in-memory 
computing alone speeds-up executions by a factor of up to 1.6, whereas 
when combined with data locality, this factor reaches 5.3. Lazy evaluation
strategies
were found to increase the likelihood of cache hits, further improving
    processing time. Such important 
speed-up values are likely to be observed on typical image processing 
operations performed on images of size larger than 75GB. A ballpark 
speculation from our model showed that in-memory computing alone will
not speed-up current 
functional MRI analyses unless coupled with data 
locality and processing around 280 subjects concurrently. Furthermore, we observe
    that emulating in-memory computing 
using in-memory file systems (tmpfs) does not reach the performance of an in-memory engine,
    presumably due to swapping to disk and the lack of data cleanup.
We conclude that Big Data processing strategies are 
worth developing for neuroimaging 
applications. 
\end{abstract}

\section{Introduction} 

Big Data processing engines have significantly improved the performance of Big Data
applications by diminishing the amount 
of data movement that occurs during the execution of an application. Locality-aware
scheduling, introduced by the MapReduce~\cite{dean2008mapreduce} framework, reduced the overall
costs associated to network transfer of data by scheduling tasks to the nodes located nearest to
the data. Reduction of data movement was further improved upon through in-memory computing~\cite{zaharia2016apache},
which ensures that data is maintained in memory between tasks whenever possible. To further
reduce the cost of data movement, lazy evaluation, the process of performing computations only when invoked,
was leveraged by Big Data frameworks to enable further optimizations such as regrouping of tasks and
computing only what is necessary.

Frameworks such as
MapReduce and Spark~\cite{zaharia2016apache} have become mainstream tools for data analytics, although many
others, such as Dask~\cite{rocklin2015dask}, are emerging.
Meanwhile, several scientific 
domains including bioinformatics, physics or astronomy, have entered 
the Big Data era due to increasing data volumes and variety. 
Nevertheless, the adoption of Big Data engines for scientific data analysis 
remains limited, perhaps due to the widespread availability of 
scientific processing engines such as Pegasus~\cite{deelman2005pegasus} or
Taverna~\cite{oinn2004taverna}, and the adaptations required in Big 
Data processing engines for scientific computing. 

Scientific applications differ from typical Big Data use 
cases, which might explain the remaining gap between Big Data and 
scientific engines. While Big Data applications mostly target text 
processing (e.g. Web search, frequent pattern mining, recommender 
systems~\cite{leskovec2014mining}) implemented in consistent software 
libraries, scientific applications often involve 
binary data such as images and signals, processed by a sequence of 
command-line/containerized tools
using a mix of programming languages (C, Fortran, Python, shell 
scripts), referred to as workflows or pipelines. With respect to infrastructure, 
Big Data applications commonly run on 
clouds or dedicated commodity clusters with locality-aware file systems 
such as the Hadoop Distributed File System 
(HDFS~\cite{shvachko2010hadoop}), whereas scientific applications are 
usually deployed on large, shared clusters where data is transferred between
data and compute nodes through shared file systems such 
as Lustre~\cite{schwan2003lustre}. Such differences in applications and 
infrastructure have important consequences. To 
mention only one, in-memory computing requires instrumentation to be 
applied to command-line tools. 

Technological advances of the past decade, in particular page caching 
in the Linux kernel~\cite{love2010linux}, in-memory file systems 
(tmpfs) and memory-mapped files might also 
explain the lack of adoption of Big Data engines for scientific 
applications. In such configurations, in-memory computing would be a feature 
provided by 
the operating system rather than by the engine itself. The frontier 
between these two components is blurred and needs to be clarified.


Our primary field of interest, Neuroimaging, is no exception to the 
generalized rise of data volumes in science due to the joint increase 
of image resolution and subject cohort sizes~\cite{van2014human}. 
Processing engines have been developed with neuroinformatics 
applications in mind, for instance Nipype~\cite{gorgolewski2011nipype} 
or the Pipeline System for Octave and Matlab 
(PSOM~\cite{bellec2012pipeline}). Big Data engines have also been used for 
neuroimaging applications, including the Thunder 
project~\cite{freeman2014mapping} and in more specific works such 
as~\cite{makkie2019fast}. However, no quantitative performance 
evaluation has been conducted on neuroimaging applications to assess the 
added-value of Big Data engines compared to traditional processing engines.

This paper addresses the following questions:
\begin{enumerate}
\item What is the effect of in-memory computing, lazy evaluation and data locality on current neuroimaging applications?
\item Can in-memory computing be effectively enabled by the operating system rather than the data processing engine?
\end{enumerate}

Answers to these questions have important implications. 
In~\cite{mehta2017comparative}, a comparative study of Dask, Spark, 
TensorFlow, MyriaDB, and SciDB on neuroinformatics use-cases is 
presented. It concludes that these systems need to be extended to 
better address scientific code integration, data partitioning, data 
formats and system tuning. We argue 
that such efforts should only be conducted if substantial performance 
improvements are expected from in-memory computing, lazy 
evaluation or data 
locality. On the other hand, neuroimaging data processing engines are 
still being developed, and the question remains whether these 
projects should just be migrated to Spark, Dask, or other Big Data 
engines.

Our study focuses on performance. We intentionally do not 
compare Big Data and scientific data processing engines on the grounds 
of workflow language expressivity, fault-tolerance, provenance capture 
and representation, portability or reproducibility, which are otherwise 
critical concerns, addressed for instance in~\cite{samba}. Besides, our 
study of performance focuses on the impact of data writes and 
transfers. It purposely leaves out task scheduling to computing 
resources, to focus on the understanding of 
data writes and movement. Task scheduling will be part of our 
discussion, however.

In terms of infrastructure, we focus on the case of High-Performance Computing 
(HPC) clusters that are typically available through University 
facilities or national computing infrastructures such as 
\href{xsede.org}{XSEDE}, \href{http://computecanada.ca}{Compute Canada} 
or \href{http://www.prace-ri.eu}{PRACE}, as neuroscientists typically use such platforms.
 We assume that HPC systems are 
multi-tenant, that compute nodes are accessible through a batch 
scheduler, and that a file system shared among the compute nodes is 
available. We intentionally did not consider distributed, 
HDFS-like file systems, as initiatives to deploy them in HPC 
centers, for instance Hadoop on-demand~\cite{krishnan2011myhadoop}, have not 
become mainstream yet.

Our methods, including performance models, processing engines, 
applications and infrastructure used, are described in 
Section~\ref{sec:methods}. Section~\ref{sec:results} presents our 
results which we discuss in Section~\ref{sec:discussion} along with the 
two research questions mentioned previously. 
Section~\ref{sec:conclusion} concludes on the relevance of Big Data 
processing strategies for neuroimaging applications.

\section{Materials and Methods} 
\label{sec:methods}

The application pipelines, benchmarks, performance data, and analysis scripts used 
to implement the methods described hereafter are all available at 
\url{https://github.com/big-data-lab-team/paper-in-mem-locality} for 
further inspection and reproducibility. Links to the processing engines 
and processed data are provided in the text.

\subsection{Engines} 

\subsubsection{Apache Spark}

Apache Spark is a well-established Scala-based processing framework for Big Data, with
APIs in Java, Python and R. Its 
generalized nature allows Spark to not only be applied to batch workflows,
but also SQL queries, iterative machine learning applications, 
data streaming and graph processing. Spark's
main features include data locality, in-memory processing and lazy evaluation,
which it achieves through its principle abstraction, the Resilient Distributed 
Dataset (RDD). 

An RDD is an immutable parallel data structure that achieves fault-tolerance 
through the concept of lineage~\cite{zaharia2010spark}. Rather than permitting
fine-grained transformations, only coarse-grained transformations (applying to
many elements in the RDD), can be applied, thereby making it simple to maintain a 
log of data modifications. This log, known as the lineage, is used
to reproduce any lost data modifications.

Two types of operations can be performed on RDDs:
transformations and actions. Applying a transformation to an RDD produces a new
child RDD through a narrow or wide dependency. A narrow dependency signifies 
that the child is only dependent on a single parent partition, whereas a child 
RDD is dependent on all parent partitions in a wide dependency. Examples of 
transformations include map, filter and join. To materialize an RDD, an
action must be performed, such as a reduce or a collect. Lazy evaluation is 
represented in Spark through the use of transformations and actions. A series of
transformations may be defined without the data ever being materialized. Using 
this strategy, Spark can optimize data processing
throughout the application.

All actions and wide dependencies require a shuffle -- Spark's most costly
operation. Every shuffle begins with each map task saving its data to local
files for fault tolerance. The shuffle operation then 
redistributes the data across the partitions as requested. A shuffle marks a 
stage boundary in Spark, where reduce-like operations will not begin 
until all dependent map tasks have completed.

Although Spark uses in-memory computing, it is not necessary for all the data to
fit in memory. Spark will spill any RDD elements that cannot be maintained in memory to disk.
Moreover, as Spark transformations generate new RDDs and numerous 
transformations may occur within a single application, Spark implements a
Least-Recently Used (LRU) eviction
policy. If an evicted RDD needs to be reused, Spark will recompute it using the
lineage data collected. As demonstrated in~\cite{freeman2014mapping}, caching
significantly improves processing times of iterative algorithms where 
RDDs are reused. It can be of even greater importance if the RDD is costly to 
recompute.

Data locality in Spark is achieved through the scheduling of tasks to partitions 
which have the
data loaded in memory. If the data is instead stored on HDFS, the scheduler will
assign it to one of the preferred locations specified by HDFS. Spark's scheduler
utilizes delay scheduling to optimize fairness and locality for all tasks.

Three different types of schedulers are compatible with Spark: 1) Spark 
Standalone, 2) YARN~\cite{vavilapalli2013apache} and 3) Mesos~\cite{hindman2011mesos}. The Spark Standalone
scheduler is the default scheduler. The YARN scheduler designed for Hadoop~\cite{white2012hadoop}
clusters and is prepackaged with Hadoop installations, whereas in contrast, 
Mesos was designed to be used in multi-tenant cluster environments. In our experiments,
we focus on the Spark Standalone cluster.

Executing Spark applications on HPC systems with Spark-unaware schedulers may
be inefficient. The amount of resources requested by Spark may impede 
Spark-cluster scheduling time. Using pilot-scheduling strategies to add 
nodes to the Spark cluster as they are allocated by the underlying 
schedulers may speedup allocation and overall processing time~\cite{paraskevakos2018pilot}. This, however, is not studied in the 
current paper.

As Spark is frequently used by the scientific community, we designed 
our experiments using the PySpark API. This is at a cost to performance 
as the PySpark code must undergo Python to Java serialization. We used Spark
version 2.3.2 installed from \url{https://spark.apache.org}.

\subsubsection{Nipype}

Nipype is a popular Python neuroimaging processing engine. It aims at being a 
solution for easily creating reproducible neuroimaging workflows. Although Nipype
does not employ any Big Data processing strategies, it provides plugins to numerous
schedulers found in most clusters readily available to researchers, 
such as the Sun/Oracle Grid Engine (SGE/OGE), \href{http://www.adaptivecomputing.com/products/torque/}{TORQUE},
\href{https://slurm.schedmd.com/}{Slurm} and \href{https://research.cs.wisc.edu/htcondor/}{HTCondor}. It also includes its own scheduler, MultiProc,
for parallel processing on single nodes. Furthermore, Nipype provides many built-in 
interfaces to commonly used neuroimaging tools that can be incorporated within the 
workflows. Nipype's ability to easily parallelize workflows in researcher-available 
cluster setups, capture detailed provenance information necessary for reproducibility,
and allow users to easily integrate existing neuroimaging tools, make it preferable
over existing Big Data solutions, which would necessitate modifications to achieve this.

Jobs, or Interfaces in Nipype, are encapsulated by a Node. A Node dictates that
the job will execute on a single input. However, the MapNode, 
a child variant of the Node, can execute on multiple inputs. All tasks in
Nipype execute in their own uniquely named subdirectory which facilitates provenance 
tracking of inputs and outputs and also enables checkpointing of the workflow.
In the case of Node failure or application modification, only the nodes which have
been modified (verified by hash) or have not successfully completed, are re-executed.

In order for Nipype to operate as intended, a filesystem shared by all the nodes 
is required. However, it is still possible to save to a non-shared 
local filesystem, but it may come at the expense of fault-tolerance, as data 
located on failed nodes will be permanently lost. Moreover, the 
user will need to ensure that the files are appropriately directed to 
the nodes that require them as there is no guarantee of data locality 
in Nipype.

We used Nipype version 1.1.4 installed through the pip package manager.

\subsection{Data Storage Locations}

Data storage location is critical to the performance of Big Data 
applications on HPC clusters.
Data may reside in the engine memory, on 
a file system whose contents reside in virtual memory (for instance 
tmpfs), on disks local to the processing node, or on a shared 
file system. Table~\ref{table:features} summarizes the Big Data 
strategies that can be used depending on the data location. In 
addition, lazy evaluation is available in Spark regardless of data 
location. The remainder of this Section explains this Table and 
provides related performance models.
\begin{table}
\centering
\begin{tabular}{c|cc}
   \rowcolor{headcolor}
    Data Location                 & In-Memory     & Data Locality        \\
    \rowcolor{headcolor}
                                  & Computing     &                     \\
                                  \hline          
In-memory                         & \cellcolor{green!25} Yes           & \cellcolor{green!25}Yes                      \\
tmpfs                             & \cellcolor{green!25} Yes           & \cellcolor{green!25}Yes                  \\
Local Disk                        & \cellcolor{orange!25}Page Caching  & \cellcolor{green!25}Yes                  \\
Shared File System                & \cellcolor{orange!25}Page Caching  & \cellcolor{red!25}No                
\end{tabular}
\setlength{\belowcaptionskip}{-10pt}
\caption{Big Data strategies on a shared HPC cluster.}
\label{table:features}
\end{table}

\subsubsection{In-Engine-Memory and In-Memory File System} 

 The main difference between storing the data in the engine memory, as in Spark, 
 and simply writing to an in-memory file system, such as tmpfs, is 
 what happens when the processed data fills up the available 
 memory. When engine memory is used, the engine must cleanup 
 unused data to avoid crashes. When an in-memory 
 file system is used, the user is responsible ensuring that the filesystem does not
 reach capacity. Should it be necessary to free up additional memory, the kernel will swap filesystem memory to disk
 and the performance will become that of local disk writes. In our experiments, 
 we will explore the configuration where the data consumed by the 
 application approaches the threshold of available memory.

\subsubsection{Local Disk} 


Storing data on local disks inevitably enables data locality, since data transfers are
not necessary when tasks are executed on the nodes where the data resides.
However, in absence of a more specific 
filesystem such as HDFS to handle 
file replication across computing nodes, data locality comes at the price
of stringent scheduling restrictions, as tasks can only be scheduled to the
single node that contains their input data.

The performance of local disk accesses is strongly dependent on the 
page caching mechanism provided by the Linux kernel, described in    
details in~\cite{love2010linux}. To summarize, data read from disk 
remains cached in memory until evicted by an LRU (Least Recently Used) 
strategy. When a process invokes the \texttt{read()} system call, the 
kernel will return the data directly from memory if the requested data 
lies in the page cache, realizing a cache \emph{hit}. Cache hits drastically speed-up data 
reads, by masking the disk latency and bandwidth behind a 
memory buffer. In effect, page caching provides in-memory computing 
transparently to the processing engine. However, page cache eviction 
strategies currently cannot be controlled by the application, which 
prevents processing engines from anticipating reads by preloading the 
cache. Scheduling strategies might be designed 
to maximize cache hits, however. For instance, lazy 
evaluation could result in more cache hits by scheduling data-dependent 
tasks on the same node.

Page caching has a more dramatic effect on disk writes, reducing their 
duration by several orders of magnitude. When a process calls the 
\texttt{write()} system call, data is copied to a memory cache that is 
asynchronously written to disk by flusher threads, when memory shrinks, when
``dirty" (unwritten) data grows, or when a 
process invokes the \texttt{sync()} system call. 
This asynchronous flushing of the page cache is called 
\emph{writeback}.

Page caching is essentially a way to 
emulate in-memory computing at the kernel level, without requiring a 
dedicated engine. The size of the page cache, however, becomes a 
limitation when processes write faster than the disk bandwidth. When 
this happens, the page cache rapidly fills up and writes are limited by 
the disk write bandwidth as if no page cache was involved.

We introduce the following basic model to describe the filling and 
flushing of the page cache by an application:
$$
d(t) = \left( \frac{D}{C} - \frac{\delta}{\gamma} \right)t + d_0,
$$
where:
\begin{itemize}
\item $d(t)$ is the amount of data in the page cache at time t
\item $D$ is the total amount of data written by the application
\item $C$ is the total CPU time of the application
\item $\delta$ is the disk bandwidth
\item $\gamma$ is the max number of concurrent processes on a node
\item $d_0$ is the amount of data in the page cache at time $t_0$
\end{itemize}

This model applies to parallel applications assuming that (1) 
concurrent processes all write the same amount of data, (2) 
concurrent processes all consume the same CPU time, (3) data is written 
uniformly along task execution. With these assumptions, all the 
processes will write at the same rate, which explains why the model 
does not depend on the total number of concurrent processes in the 
application, but only on the max number of concurrent processes 
executing on the same node ($\gamma$). While these 
assumptions would usually be violated in practice, this simple 
model already provides interesting insights on the performance of disk 
writes, as shown later. Naturally, the model also ignores other 
processes that might be writing to disk concurrently to the 
application, which we assume negligible here. 

In general, an application should ensure that $\dot d$ remains negative 
or null, leading to the following inequality:
\begin{equation}
\frac{D}{C} \leq \frac{\delta}{\gamma} \label{eq:page-cache-inequality}
\end{equation}
This defines a D/C (data-write) rate beyond which the page cache becomes 
asymptotically useless. It should be noted that the transient phase 
during which the page cache fills up might last a significant amount of time, 
in particular when $\dot d$ is positive and small. We intentionally do not model the 
transient phase as it requires detailed knowledge of difficult to estimate parameters  
such as the page cache size and the initial amount 
of data in it ($d_0$).

 We will use Equation~\ref{eq:page-cache-inequality} to 
define our benchmarks and interpret the results. It 
should be noted that leveraging the page cache, and therefore ensuring 
that Equation~\ref{eq:page-cache-inequality} holds, has important 
performance implications: with page caching, the write throughput will 
be that of memory, while without page caching it will be that of the 
disk.


\subsubsection{Shared File System}

We model a shared file system using its global realized bandwidth 
$\Delta$, shared by all concurrent processes in the cluster. We are 
aware that such a simplistic model does not describe at all the 
intricacies of systems such as Lustre. In particular, metadata 
management, RPC protocol optimizations and storage optimizations are 
all covered under the realized bandwidth. We do, however, consider the 
effect of page caching in shared file systems, since in Linux 
writes to network-mounted volumes benefit from this feature too.

As in the local disk model, we note that page caching will only be 
useful when the flush bandwidth is greater than the write throughput of 
the application, that is:
\begin{equation}
\frac{D}{C} \leq \frac{\Delta}{\Gamma}, \label{eq:page-cache-sharedfs}
\end{equation}
where $\Gamma$ is the max number of concurrent processes \emph{in the cluster}. 
Note that $\frac{\Delta}{\Gamma}$ will usually be much lower than 
$\frac{\delta}{\gamma}$.






\subsection{Infrastructure} 

 All experiments were executed on 
 \href{https://www.dellemc.com/resources/en-us/asset/sales-documents/products/storage/h16221-hpc-lab-brochure.pdf}{Dell 
 EMC's Zenith cluster}, a Top-500 machine in the 
 \href{https://www.dellemc.com/en-us/solutions/high-performance-computing/HPC-AI-Innovation-Lab.htm}{Dell 
 EMC HPC and AI Innovation Lab}, running Slurm. For the Spark 
 experiments, a Spark cluster was started on a
 Slurm allocation comprised of 16 dedicated nodes.
 Each Compute node has Red Hat Enterprise Linux Server release 7.4 (Maipo) 
 as the base operating system with kernel version 3.10.0-693.17.1.el7.x86\_64
 (patched for Spectre/Meltdown). Dell EMC PowerEdge C6420 with dual Intel Xeon
 Gold 6148/F processors (40 cores per node) and 192GB ($12\times16$~GB), 2666 MHz
 memory, serve as the compute nodes. Each compute has a 120GB M.2 SATA SSD as 
 local disk. A Dell HPC Lustre Solution with a raw storage of 960TB is 
 accesible on each compute node through a 100~Gb/s Intel OmniPath network. All
 the nodes connect to a director switch in a 1:1 non-blocking topology.
 The realized write bandwidth of the local disk, 
 Lustre file system and tmpfs were measured by sequentially writing various numbers
 of image blocks containing random intensities, to avoid 
 caching effects (see: \href{https://github.com/big-data-lab-team/paper-in-mem-locality/blob/master/benchmark_scripts/measure_bandwidth.py}{measure\_bandwidth.py}).
 They are reported in Table~\ref{table:bdwdths}.

\begin{table}
\centering
\begin{tabular}{c|c}
\rowcolor{headcolor}
Data location & Measured write bandwidths (MB/s)\\
\hline
tmpfs                 & 1377.18 \\
Local disk ($\delta$) & 193.64  \\
Lustre   ($\Delta$)   & 504.03 \\
\end{tabular}
\setlength{\belowcaptionskip}{-10pt}
\caption{Measured bandwidths}
\label{table:bdwdths}
\end{table}

\subsection{Datasets} 

We used BigBrain~\cite{amunts2013bigbrain}, a 75GB 40$\mu$m isotropic 
histological image of a 65-year-old human brain. The BigBrain was selected due 
to its uniqueness, as there does not yet exist a
higher-resolution image of a human brain. Moreover, there currently exists a 
lack of standardized tools for processing the BigBrain as a consequence 
of its size. To examine the effects processing the BigBrain has on Big Data 
strategies, we partitioned the full $3845\times3015\times3470$ voxel image into 30 ($5\times3\times2$)
chunks, 125 ($5\times5\times5$) chunks and 750 ($5\times15\times10$) chunks. Additionally, the full 
image was also split in half ($769\times603\times347$ voxels) and the half image was 
partitioned into 125 chunks.

Processing large images is only considered to be part of the Big Data problem 
in neuroscience. The other problem being the processing large MRI datasets, 
that is, datasets consisting of many small brain images belonging to 
various different subjects. This situation is commonly observed in 
functional MRI (fMRI), where it is becoming increasingly common to 
process data from hundreds of subjects. Although we have not explored 
explicitly the processing of large MRI datasets, the 75GB BigBrain is 
within the size ballpark~\cite{van2014human} of 
MRI datasets commonly processed in today's studies.

Since both small and large datasets may need to be processed using
the same analysis pipeline, we 
examined the effects of the data management strategies on small 
data as well. For this, we selected a 12MB T1W image belonging to 
subject 1 of OpenNeuro's 
\href{https://openneuro.org/datasets/ds000001/versions/00006/file-display/sub-01:anat:sub-01_T1w.nii.gz}{ds000001 
dataset version 6}. In order to split the image into 125 equal-sized chunks, it 
was necessary to zero-pad the image to the dimensions
$165\times200\times200$ voxels, which subsequently increased the total image size to 13MB.

\subsection{Applications} 
\begin{algorithm}\caption{Incrementation}\label{alg:incrementation}
    \begin{algorithmic}[1]
    \Input
    \Desc{$x$}{a sleep delay in seconds}
    \Desc{$n$}{a number of iterations}
    \Desc{$C$}{a set of image chunks}
    \Desc{$fs$}{filesystem to write to (mem, tmpfs, local disk, Lustre).}
    \EndInput
    \ForEach{$chunk \in C$}
    \State read $chunk$ from Lustre
    \For{$i \in [1, n]$}
        \State $chunk\gets chunk+1$
        \State sleep $x$
        \If{$i < n$}
        \State save $chunk$ to $fs$
        \EndIf
    \EndFor
    \State save $chunk$ to Lustre
    \EndFor
\end{algorithmic}
\end{algorithm}

To effectively investigate how the different strategies impact processing, we 
selected a simple incrementation pipeline (Algorithm~\ref{alg:incrementation}) 
that consists exclusively of map 
stages. A series of map-only stages would enable us to evaluate the effects of
in-memory computing when data locality is preserved. Incrementation was 
selected 
over other applications, such as binarization, as it ensured that a new image
was created at each step (i.e. no caching effects within the executing 
application). Each partitioned chunk was incremented by 1, in parallel, by
each task. As incrementing images is not a time consuming process,
we added a sleep delay to the tasks to study the effects 
of tasks duration. The incremented chunks would be either 
maintained in-memory (Spark only) or saved to either tmpfs, local disk or 
Lustre (Spark and Nipype). Should more than a single iteration be requested, 
the incremented chunks would be incremented again and saved to the same file system. 
This would repeat until the number of requested iterations had elapsed. In all 
conditions, the first input chunks and final output chunks would reside on Lustre. 
We chose to perform our initial input/final output on Lustre as local storage is
typically only accessible to a user for the duration of the execution in HPC
environments.

\subsection{Experiments}

We conducted four experiments in which we varied (1) the number of 
iterations in the application ($n$ in 
Algorithm~\ref{alg:incrementation}), (2) the task duration ($x$), (3) 
the chunk size given a constant image size, (4) the total image size. To 
evaluate the page-cache model, experiment conditions fell in different 
regions of Equations~\ref{eq:page-cache-inequality} 
and~\ref{eq:page-cache-sharedfs}, as summarized in 
Table~\ref{table:experiments}. Among the 16 nodes available, 1 was 
dedicated to the Spark master and driver, and the remaining 15 were 
used as compute nodes. Since data locality is not normally preserved in 
Nipype (a new Slurm allocation is requested for each processed task), 
we instrumented Nipype to ensure data locality (see: \href{https://github.com/big-data-lab-team/paper-in-mem-locality/blob/master/benchmark\_scripts/run\_benchmarks.py}{run\_benchmarks.py}).
That is, the chunks were split into partitions, and for each partition,
we requested a Slurm allocation to process the entire 
pipeline in parallel using Nipype's MultiProc scheduler on a given 
node. This was possible as no communication was required between the 
processed chunks.

\begin{table*}
\centering
\begin{tabular}{c|ccc|cccc|cc}
  \rowcolor{headcolor}
  \multicolumn{10}{c}{Experiment 1: Number of Iterations}\\
  \hline
  \rowcolor{headcolor}
 n &
 D (GB) & C (s) & D/C (MB/s) &
 $\gamma$ & $\delta/\gamma$ (MB/s) & $\Gamma$ & $\Delta/\Gamma$ (MB/s)&
 (D/C)/($\delta/\gamma$) & (D/C)/($\Delta/\Gamma$)\\
 \hline
1   &75    & 430    & 178.6 & 9 & 21.5 & 125 & 4.0 & \cellcolor{red!20} 8.3 & \cellcolor{red!20} 44.7\\
10  & 750   & 4,300 & 178.6 & 9 & 21.5 & 125 & 4.0 & \cellcolor{red!20} 8.3 & \cellcolor{red!20} 44.7\\
100 & 7,500 & 43,000& 178.6 & 9 & 21.5 & 125 & 4.0 & \cellcolor{red!20} 8.3 & \cellcolor{red!20} 44.7\\
\hline
  \multicolumn{10}{c}{}\\

\rowcolor{headcolor}
  \multicolumn{10}{c}{Experiment 2: Task Duration}\\
  \hline
  \rowcolor{headcolor}
 x (s) &
 D (GB) & C (s) & D/C (MB/s) &
 $\gamma$ & $\delta/\gamma$ (MB/s) & $\Gamma$ & $\Delta/\Gamma$ (MB/s)&
 (D/C)/($\delta/\gamma$) & (D/C)/($\Delta/\Gamma$)\\
 \hline
 2.4  & 750 & 3,000   & 256   & 9 & 21.5 & 125 & 4.0 &  \cellcolor{red!20} 11.9 & \cellcolor{red!20} 64   \\
 3.44 & 750 & 4,300   & 178.6 & 9 & 21.5 & 125 & 4.0 &  \cellcolor{red!20} 8.3 & \cellcolor{red!20} 44.7   \\
 7.68 & 750 & 9,600   & 80    & 9 & 21.5 & 125 & 4.0 &  \cellcolor{red!20} 3.7 & \cellcolor{red!20} 20   \\
 320  & 750 & 400,000 & 1.9   & 9 & 21.5 & 125 & 4.0 &  \cellcolor{green!20} 0.09 & \cellcolor{green!20} 0.48  \\
 \hline
  \multicolumn{10}{c}{}\\
  
 \rowcolor{headcolor}
  \multicolumn{10}{c}{Experiment 3: Number of Chunks}\\
  \hline
  \rowcolor{headcolor}
 chunks &
 D (GB) & C (s) & D/C (MB/s) &
 $\gamma$ & $\delta/\gamma$ (MB/s) & $\Gamma$ & $\Delta/\Gamma$ (MB/s)&
 (D/C)/($\delta/\gamma$) & (D/C)/($\Delta/\Gamma$)\\
 \hline
 30  & 750 & 4,400   & 174.6 & 2  & 96.8 & 30  & 16.8 &  \cellcolor{red!20} 1.8 & \cellcolor{red!20} 10.4   \\
 125 & 750 & 4,400   & 174.6 & 9  & 21.5 & 125 & 4.0 &  \cellcolor{red!20} 8.1 & \cellcolor{red!20} 43.7  \\
 750 & 750 & 4,400   & 174.6 & 25 & 7.7  & 375 & 1.3 &  \cellcolor{red!20} 22.7 & \cellcolor{red!20} 134.3  \\

\hline
  \multicolumn{10}{c}{}\\ 
 \rowcolor{headcolor}
  \multicolumn{10}{c}{Experiment 4: Image Size}\\
  \hline
  \rowcolor{headcolor}
 image  &
 D (GB) & C (s) & D/C (MB/s) &
 $\gamma$ & $\delta/\gamma$ (MB/s) & $\Gamma$ & $\Delta/\Gamma$ (MB/s)&
(D/C)/($\delta/\gamma$) & (D/C)/($\Delta/\Gamma$)\\
 \hline
 BigBrain      & 750   & 2,200   & 349.1     & 9  & 21.5  & 125 & 4.0 &  \cellcolor{red!20} 16.2 & \cellcolor{red!20} 87.3   \\
 Half BigBrain & 375   & 2,200   & 174.6     & 9  & 21.5  & 125 & 4.0 &  \cellcolor{red!20} 8.1 & \cellcolor{red!20} 43.7   \\
 MRI           & 0.127 & 2,200   & 0.06      & 9  & 21.5  & 125 & 4.0 &  \cellcolor{green!20} 0.003 & \cellcolor{green!20} 0.015  
\end{tabular}
\setlength{\belowcaptionskip}{-10pt}
\caption{Experiment conditions. Red cells denote the conditions where 
the inequalities in Equations~\ref{eq:page-cache-inequality} 
and~\ref{eq:page-cache-sharedfs} do not hold, i.e., the page cache is 
asymptotically useless. Green cells show the conditions where the page cache covers all data writes.}
\label{table:experiments}
\end{table*}

For our first incrementation experiment, we investigated the effects of 
Big Data strategies on varying total data size. To achieve this, we 
increased the number of incrementation iterations from 1, 10 and 100 
times. The total data size would then increase from 75GB, at 1 
iteration, to 7,500GB, at 100 iterations. The total number of chunks 
was 125. Chunks all ran concurrently ($\Gamma$=125) and were 
equally balanced among 15 nodes, leading to 8 or 9 concurrent jobs per 
node ($\gamma$=9). Task duration was fixed at 3.44 seconds.

In the second experiment, we evaluated the effects of Big Data 
strategies on varying task duration. If the page cache has sufficient 
time to flush, it would be expected that in-memory computing 
and local disk perform equivalently. We varied the task 
duration between 2.4 and 320 seconds such that the D/C falls into 
different regions of Equations~\ref{eq:page-cache-inequality} 
and~\ref{eq:page-cache-sharedfs}. The number of chunks was maintained 
at 125, leading to $\Gamma$=125 and $\gamma$=9. The number of iterations was
fixed to 10.

As a third incrementation experiment, we were interested in the effects of 
chunk size on Big Data strategies. Naturally, a greater chunk size signifies 
a decrease in parallelism. However, it also signifies an increase in 
sequential I/O (increased $\Delta/\Gamma$ and $\delta/\gamma$). For 
this experiment we partitioned the complete BigBrain image into 30, 125 
and 750 chunks, corresponding to $\gamma$ values of 2, 9 and 25 
respectively. While Spark attempted to load-balance the data, it used 
up only 25 of the 40 cores for 750 chunks. In contrast, Nipype tried to 
use up as many cores as possible. Unlike the previous 
experiment, the D/C rate was kept static at 178.6MB/s, however, this 
ratio ensured that different regions of the inequality were reached 
depending on amount of parallelism. The number of iterations was
fixed to 10, and the task duration was adjusted so that C remained constant at 4,400s.

For our fourth and final incrementation experiment, we investigated the effects 
of the strategies on different image sizes. We selected the 75GB BigBrain, the
38GB half BigBrain and the 13M T1W MRI image for this experiment. The number of
chunks was fixed to 125. Similarly to the previous experiment, the 
total sequential compute time was fixed (10 iterations, 1.76 seconds 
per task), however, due to varying size in total data processed (D), 
the D/C rate varied. Once again, we ensured that the D/C rate fell in 
multiple different regions of the inequality. The D/C rate ranged from 
349.1 MB/s for BigBrain and 174.6MB/s for half BigBrain, to 0.06MB/s for the MRI 
image. Only the 0.06MB/s MRI satisfied the inequality for both Lustre and local 
disk.







\section{Results} 
\label{sec:results}


\begin{figure*}
\begin{subfigure}{\columnwidth}
    \centering
    \includegraphics[width=\columnwidth]{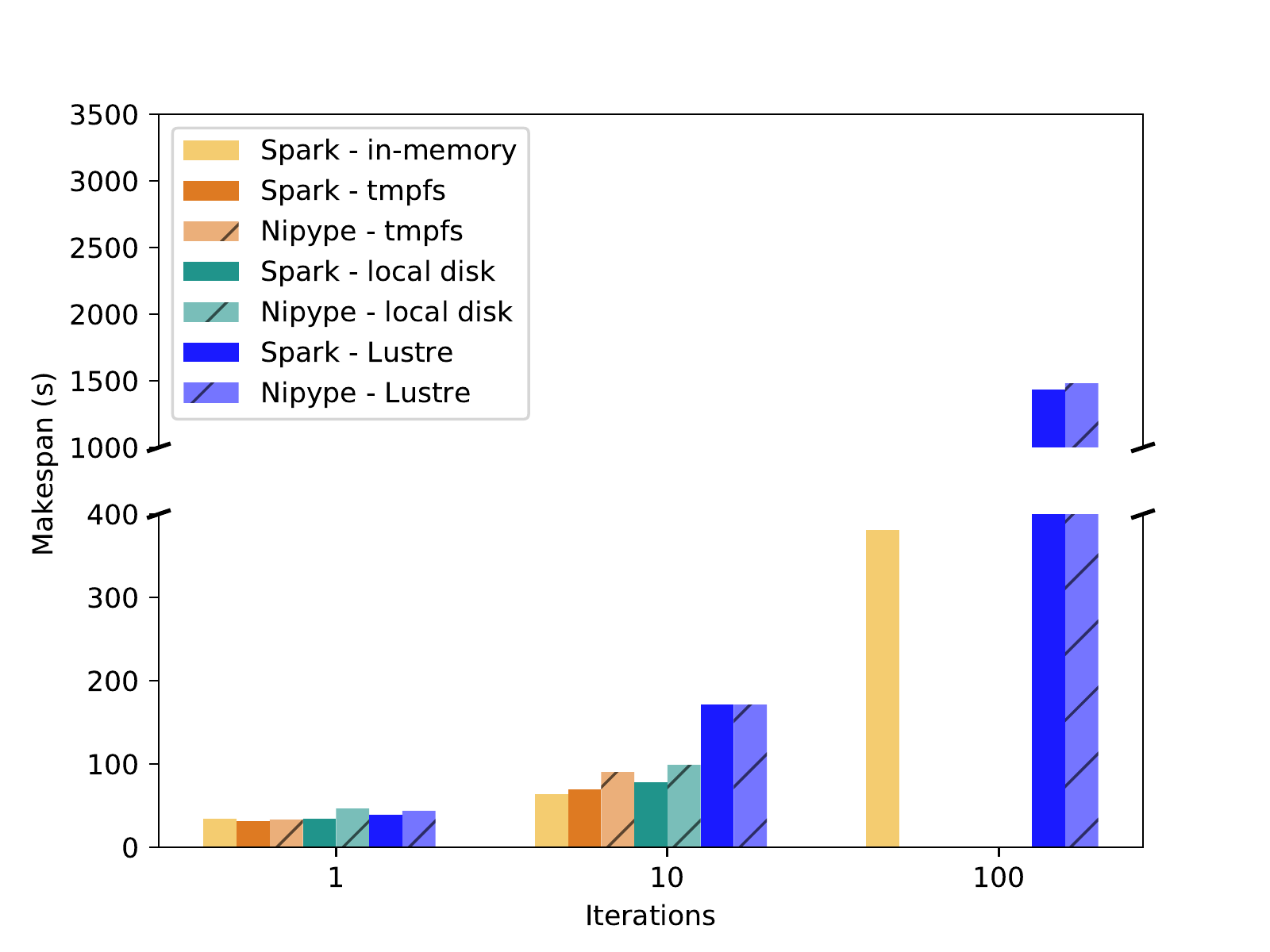}%
    \caption{Experiment 1: complete BigBrain, 125 chunks, 3.44-second tasks.}\label{fig:iterations}
\end{subfigure}
\begin{subfigure}{\columnwidth}
    \centering
    \includegraphics[width=\linewidth]{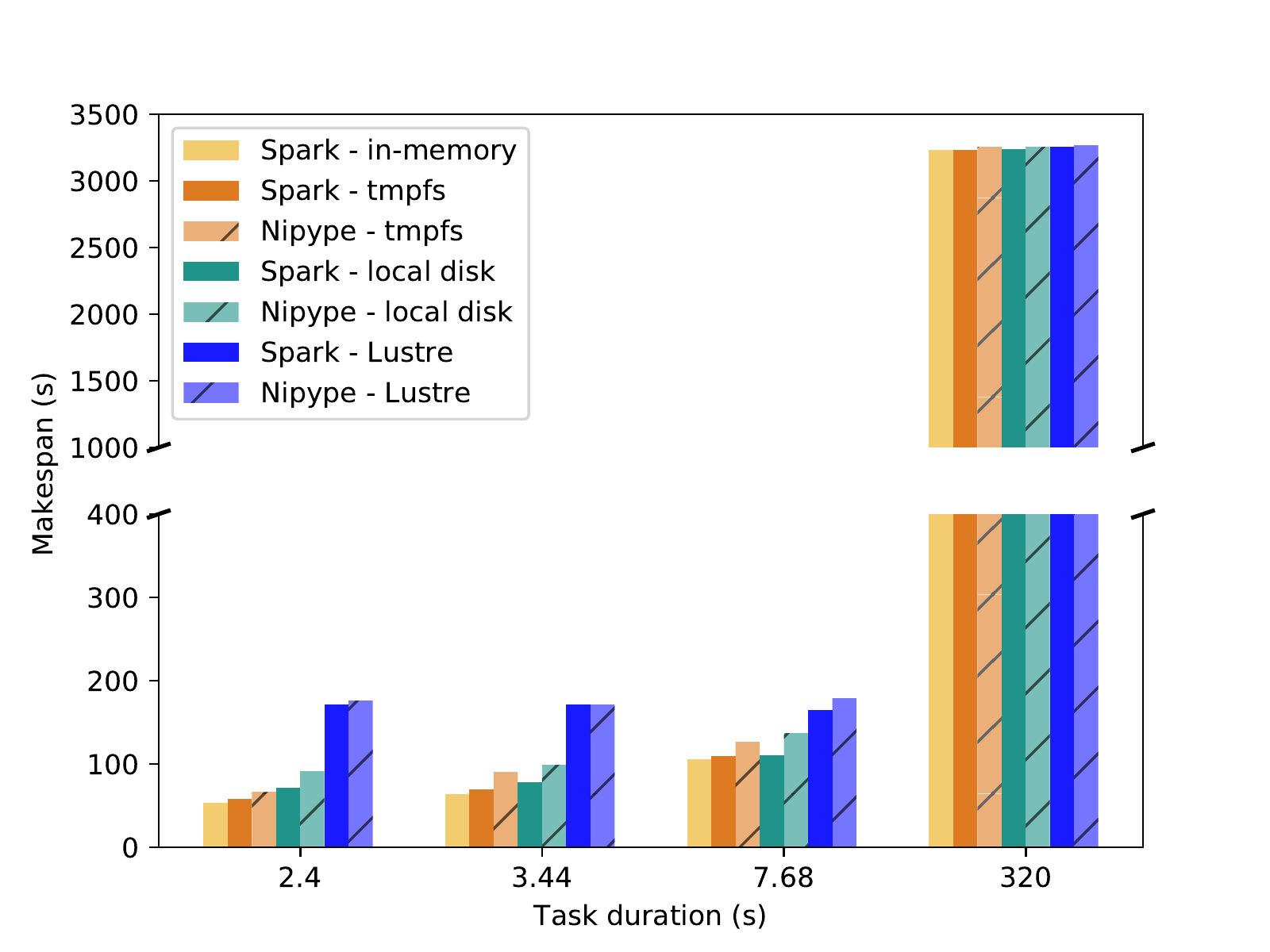}
    \caption{Experiment 2: complete BigBrain, 125 chunks, 10 iterations.}\label{fig:cputime}
\end{subfigure}
\begin{subfigure}{\columnwidth}
    \centering
    \includegraphics[width=\linewidth]{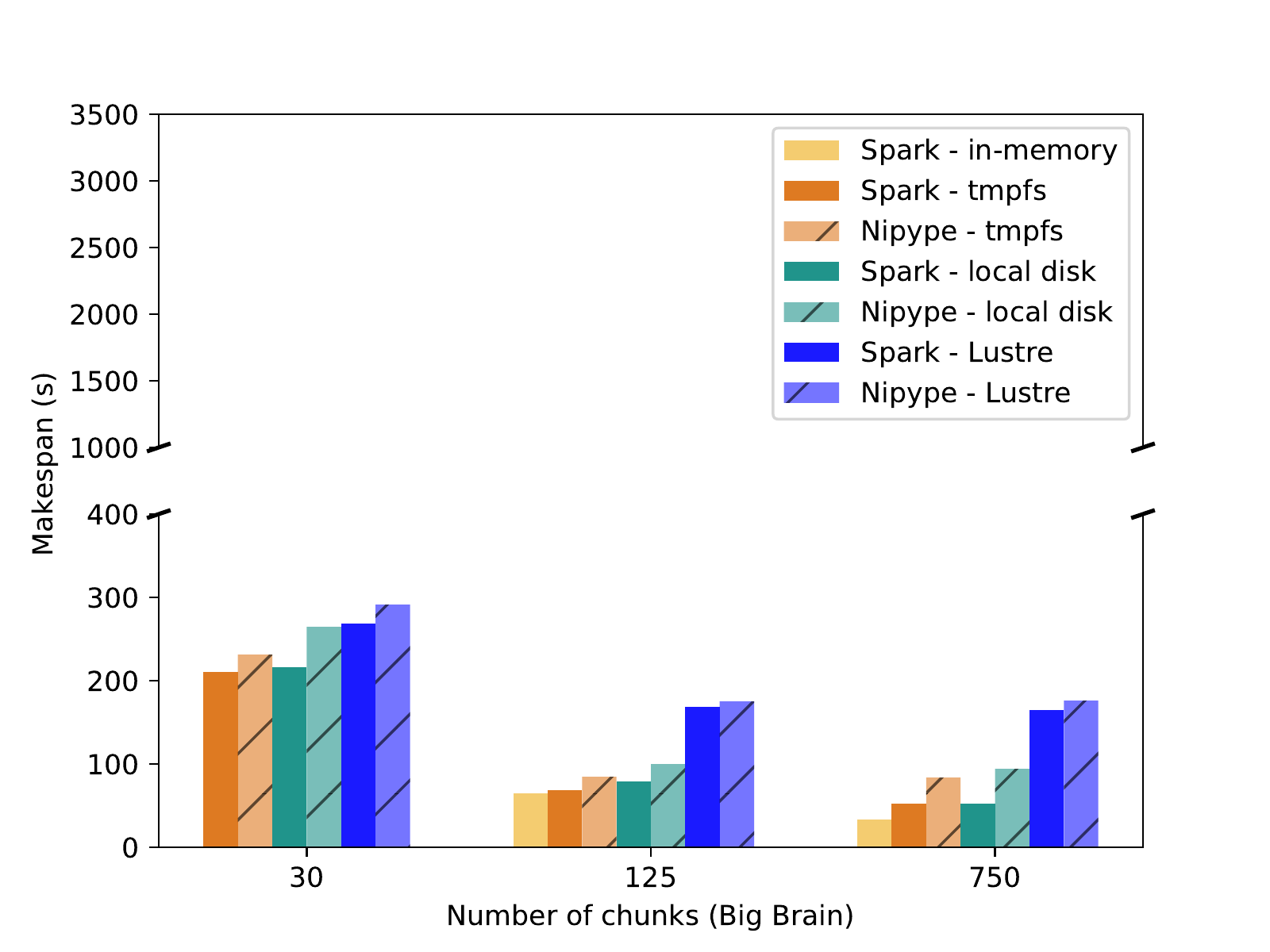}
    \caption{Experiment 3: complete BigBrain, 10 iterations, C= 4,400s.}\label{fig:numchunks}
\end{subfigure}
\begin{subfigure}{\columnwidth}
    \centering
    \includegraphics[width=\linewidth]{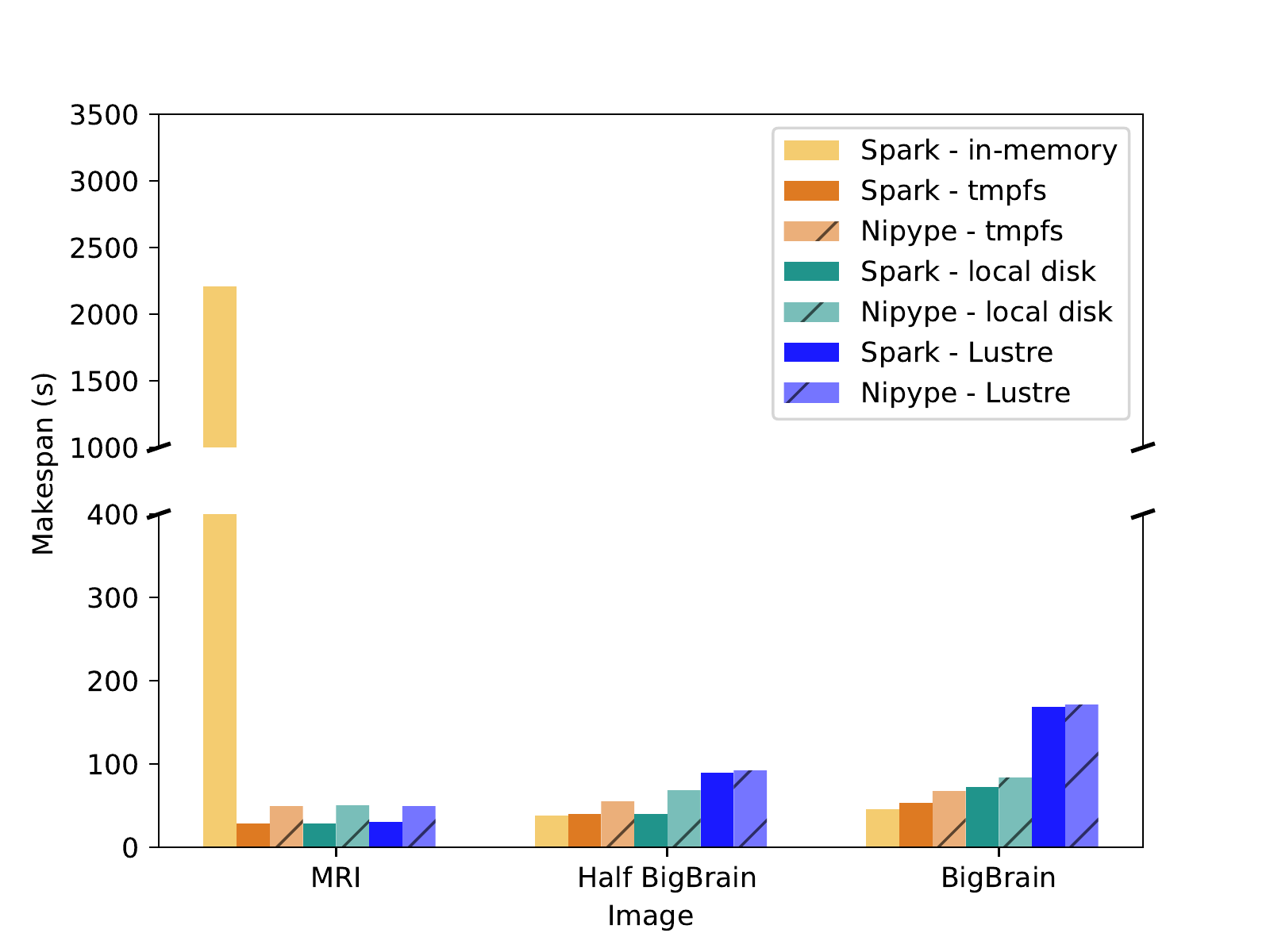}
    \caption{Experiment 4: 125 chunks, 10 iterations, 1.76-second tasks.}\label{fig:datasize}
\end{subfigure}
\setlength{\belowcaptionskip}{-10pt}
\caption{Experiment results: Makespans of Spark and Nipype writing to memory, tmpfs, local disk and Lustre.}
\label{fig:results}
\end{figure*}
\subsection{Experiment 1: Number of Iterations}

Fig.~\ref{fig:iterations} shows the difference between the different 
filesystem choices given varying number of iterations. At 1 iteration, all 
filesystems behave the same, although the application was writing faster than the
disk bandwidth. This is because application data was not saturating the page cache
(transient phase). The page cache, on Zenith, occupies 40\% of 
total memory. With 192~GB of RAM on each node, 76.8~GB of 
dirty data could be held in a node's page cache at any given time.
As the total amount of data written by the 
application increases to 750~GB, there is a greater disparity between 
Lustre and in-memory (2.67~x slower, on average). Local disk 
performance, however, 
is still comparable to memory (1.38~x slower, on average). Despite local disk and 
Lustre both being in transient state, local disk encounters less contention 
than what would be found on Lustre. 

At 100 iterations, or 7,500~GB, Lustre can be found to be, on average, 3.82~x 
slower than Spark in-memory. The 
slowdown experienced can be explained by the smaller percentage of total data 
residing in the page cache at a given time, compared to 10 iterations. 
Therefore, the effects of Lustre bandwidth are more significant in 
this application. At 100 iterations, the application was writing 500GB 
per node (7,500 GB / 15 nodes) and hence could not run on tmpfs or local disk.

While there is some variability that can be seen in Fig.~\ref{fig:iterations} 
between the two engines, this believed to be insignificant, and potentially due 
to SLURM node allocation delays in our launching of Nipype.


\subsection{Experiment 2: Task Duration}
%

Increasing task duration ensured that all file systems had a comparable performance
(Fig.~\ref{fig:cputime}). Lustre, for instance, is approximately 1.01x slower
than Spark in-memory at a task duration of 320 seconds, whereas it is 
approximately 3.25x slower that Spark in-memory with 2.4 second tasks. This 
pattern corroborated our page-cache model which postulates that 
data movement costs will have little impact on compute-intensive tasks. The 
reasoning behind this is that longer tasks give the page cache more time to flush 
between disk writes.

\subsection{Experiment 3: Image Block Size}

As can be seen in Fig.~\ref{fig:numchunks}, makespan decreases with increasing
number of chunks. This is due to the fact that parallelism 
increases with an increase in number of chunks. At 30 chunks, 
only 2 CPUs per node are actively working. At 125 chunks, this changes 
to a maximum of 9 CPUs per node, and at 750 chunks, up to 40 CPUs can be 
active.

Due to a size limitation of 2~GB imposed on Spark partitions, Spark with in-memory
computing processing 30 chunks was not performed.

Local disk and tmpfs perform comparably for all 
conditions, with Lustre being significantly slower. As with varying the number 
of iterations, Lustre is slower due to increased filesystem contention, which 
is, at minimum, 15~x greater than contention on local disk, due to the number of nodes
used. With an increase in 
number of chunks, local disk and tmpfs makespans begin to converge. A potential 
explanation for this may be that tmpfs is utilizing swap space. As concurrency 
increases, the memory footprint of the application also increases. It is 
possible that at 750 chunks, swapping to disk is required by tmpfs, thus 
resulting in similar processing times as local disk.

Swapping may also be an explanation for the variance between Spark in-memory 
and tmpfs performance. While Spark may also spill to disk, it only does so when
data does not fit in memory. As none of the RDDs generated throughout the 
pipeline were cached and all data concurrently accessed could be mantained 
in-memory, spilling to disk was not necessary.

\subsection{Experiment 4: Image Size}

Increasing overall data size decreases performance, as can be seen in 
Fig.~\ref{fig:datasize}. When the data size is very small (e.g. MRI image) all 
file system makespans are comparable. This is due to the 
fact that page cache can be leveraged fully regardless of file system. However, 
this time, Spark in-memory performed significantly worse than all other 
filesystems, with a makespan of 2,211 seconds. Upon further inspection,
it appeared that Spark in-memory executed
in a sequential order, on a single worker node. Lack of parallelism for the MRI 
image may be a result of Spark's maximum partition size, which is by default 128~MB
-- significantly larger than the 13~MB MRI image. 

At half BigBrain, the makespan differences become apparent in both local disk 
and Lustre, with Lustre becoming 2.4x slower than in-memory. This 
can be attributed to page cache saturation, as predicted by the model for both 
half the BigBrain image and the complete BigBrain. Only the MRI image was 
predicted to fall within the model constraints. 

When the complete BigBrain is processed, the disparity between the different 
filesystems becomes even greater. Lustre becomes 3.68x slower, whereas local 
disk becomes 1.68x slower. An explanation for this is that the page cache fills 
up faster due to data size.

\subsection{Page Cache Model Evaluations}
\begin{figure*}
\begin{subfigure}{0.5\linewidth}
\centering
    \includegraphics[width=\textwidth]{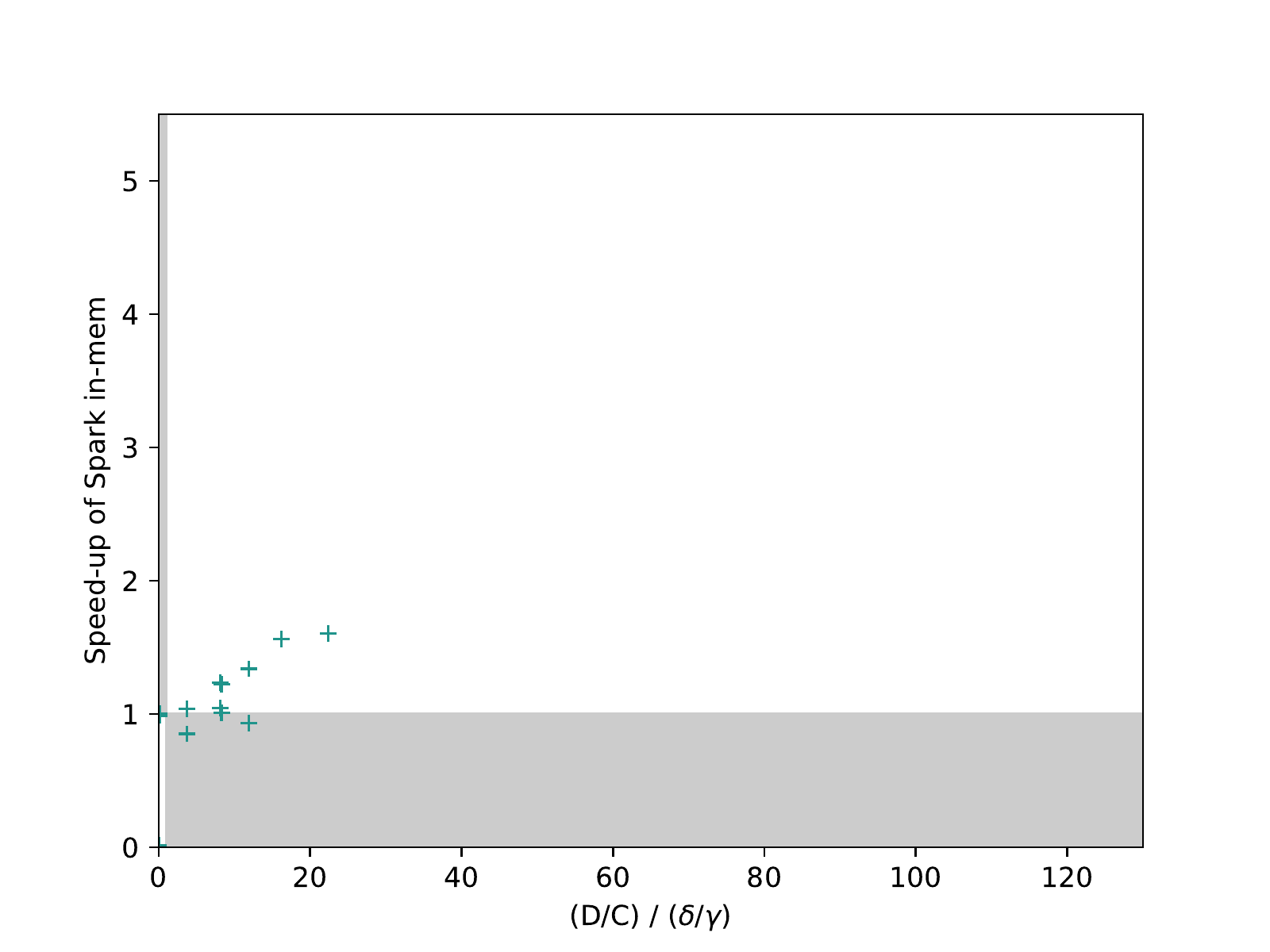}
\caption{Local Disk}
\label{fig:modeleval-local}
    \end{subfigure}%
\begin{subfigure}{0.5\linewidth}
\centering
    \includegraphics[width=\textwidth]{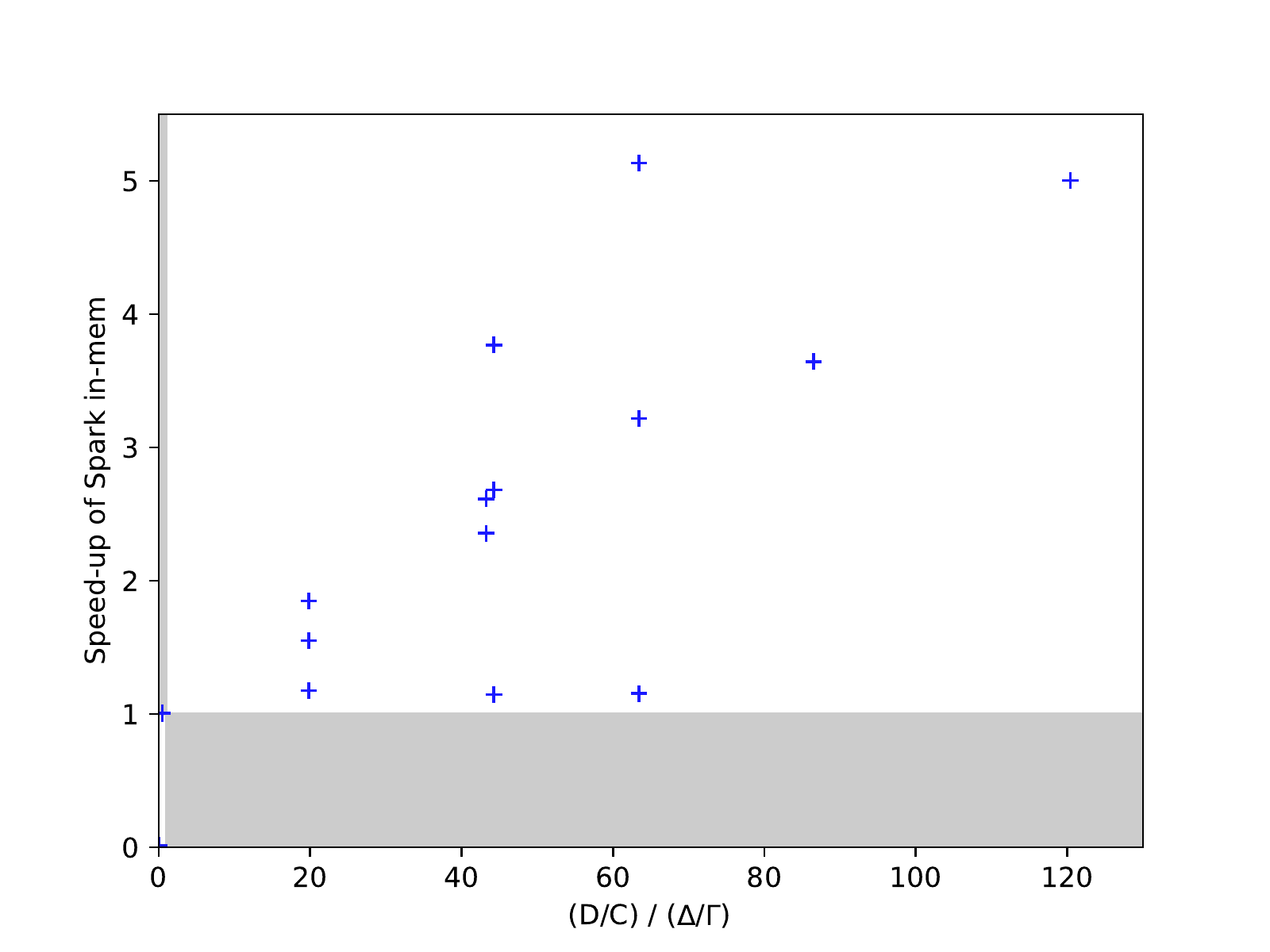}
\caption{Lustre}
\label{fig:modeleval-lustre}
\end{subfigure}
\setlength{\belowcaptionskip}{-10pt}
    \caption{Page cache model evaluation. Grey 
             regions denote areas that violate model predictions.}
\label{fig:modeleval}
\end{figure*}

In order to evaluate the page cache model, we compared the observed 
speedup ratio provided by in-memory computing to the (D/C) / 
($\delta$/$\gamma$) and (D/C) / ($\Delta$/$\Gamma$) ratios 
(Fig.~\ref{fig:modeleval}). Speed-up ratios were computed as the ratio 
between the makespan obtained with Spark on local disk or Lustre, and 
the makespan obtained with Spark for in-memory computing. Experiments 
for which there were no in-memory equivalent (i.e. BigBrain split into 
30 
chunks) 
were not considered.

Results show that, overall, the model correctly predicted the effect of 
page cache on processing times for local disk and Lustre. That is, the 
speed-up provided by in-memory computing was larger than 1 for D/C 
rates larger than $\delta/\gamma$ (local disk) or $\Delta/\Gamma$ 
(Lustre). Conversely, the speed-up provided by in-memory computing 
remained close to 1 for D/C rates smaller than $\delta/\gamma$ (local 
disk) or $\Delta/\Gamma$ (Lustre). The two points close to the origin 
correspond to the sequential processing of the MRI image by Spark 
mentioned previously.

Points which violated 
model predictions were found at 1 iteration, where page cache would not 
have been saturated in spite of a high D/C (transient state). However, 
in all cases, the ``1" boundary was never trespassed by more than a 
factor of 0.19, and is therefore likely a result of system variability.

\section{Discussion} 
\label{sec:discussion}

\subsection{Effect of In-Memory Computing}
We measured the effect of in-memory computing by comparing the runs 
of Spark in-memory (yellow bars in Fig.~\ref{fig:results}) to the ones 
of Spark on local disk (non-hatched green bars). The speed-up provided 
by in-memory computing is also reported in 
Fig.~\ref{fig:modeleval-local}. The speed-up provided by in-memory 
computing increases with (D/C) / ($\delta$/$\gamma$), as expected from 
the model. In our experiments, it peaked at 1.6, for a ratio of 16.2. 
This correspond to the processing of the BigBrain with 125 chunks and 
1.76-second tasks in experiment 4 (total computing time C=2,200s), 
which is typically encountered in common image processing tasks such as 
denoising, intensity normalization, etc. The speed up of 1.6 is also 
reached with a ratio of 22.7 in experiment 3, obtained by processing 
the BigBrain with 750 chunks.

The results also allow us to speculate on the effect of in-memory 
computing on the pre-processing of functional MRI, another typical use 
case in neuroimaging. Assuming an average processing time of 20 minutes 
per subject, which is a ballpark value commonly observed with the 
popular \href{https://www.fil.ion.ucl.ac.uk/spm/}{SPM} or \href{https://fsl.fmrib.ox.ac.uk/fsl/fslwiki/}{FSL} packages,
an input data size of 100MB per subject, 
and an output data size of 2GB (20-fold increase compared to input 
size), the D/C rate would be 1.8MB/s, which would reach the 
$\delta/\gamma$ threshold measured on this cluster for $\gamma=108$, 
that is, if 108 subjects were processed on the same node. This is very 
unlikely as the number of CPUs per node was 40. We therefore conclude 
that in-memory computing is likely to be useless for fMRI analysis.
Naturally, this estimate is strongly dependent on the characteristics of the cluster.

\subsection{Effect of Data Locality}

We measure the effect of data locality by comparing the runs of Spark
on local disk (non-hatched green bars in Fig.~\ref{fig:results}) to the 
ones of Spark on Lustre (non-hatched blue bars). The speed-up provided by 
local execution peaked at 3.2, for 750 chunks in experiment 3. Overall, 
writing locally was usually preferable over writing to Lustre, as a result of the
lower contention on local disk. Although it may be true that
network bandwidths exceed that of disks~\cite{ananthanarayanan2011disk},
locality remains important 
as contention on a shared filesystem tends to be much higher than on local disk.
The only time writing locally did not
have significant impact over Lustre was in experiments 1 and 4, at 1 iteration and 
when processing the MRI image, respectively. In both these scenarios, the Lustre 
writes did not impact performance as the data was able to be written to page cache
and flushed to Lustre asynchronously.

\subsection{Combined Effect of In-Memory and Data Locality}
We measure the combined effect of data locality and in-memory computing 
by comparing the runs of Spark in-memory (yellow bars in 
Fig.~\ref{fig:results}) to the ones of Spark on Lustre (non-hatched 
blue bars). The speed-up provided by the combined use of data locality 
and in-memory computing is also reported in 
Fig.~\ref{fig:modeleval-lustre}. The provided speed-up increases with 
(D/C) / ($\Delta$/$\Gamma$), as expected from the model. In our 
experiments, it peaked around 5, for ratios of 120.4 and 64. Again, 
this configuration is likely to happen in typical image processing 
tasks performed on the BigBrain.

As for the fMRI speculation, the D/C 
rate of 1.8MB/s would reach the $\Delta/\Gamma$ threshold for 
$\Gamma=280$, which is a realistic number of subjects to process on a 
complete cluster. Naturally, this estimate is highly dependent on the observed bandwidth
of the shared file system ($\Delta$).

\subsection{Effect of Lazy Evaluation}

The effects of lazy evaluation can be seen throughout the experiments. Nipype 
was found to be slower than Spark in most experiments. While the Nipype execution
graph is 
generated prior to workflow execution, there are no optimizations
to ensure that the least amount of work is performed to produce the required 
results. 

During the processing of Experiment 3, 750 chunks were processed in two batches
for both Spark and Nipype due to CPU limitations.
Rather than running each iteration on the full dataset, as with Nipype, 
Spark opted to perform all the iterations on the first batch (load, increment, save),
and then proceeded to process the second batch. Such an optimization is important, 
even when processing data on disk, as it would presumably increase the occurrence
of cache hits. This may partially explain the speedup seen at 750 chunks in
Figure~\ref{fig:numchunks}.

\subsection{Can tmpfs and Page Caches Emulate In-Memory Computing?}

Although tmpfs and page cache do improve performance, as seen in
Figure~\ref{fig:results}, they do not always perform equivalently to in-memory. 
Tmpfs's main limitation is that data residing on it may be swapped to disk if 
the system's memory usage is high. When it reaches this point, its performance 
slows down to swap disk bandwidth, as observed in Figure~\ref{fig:datasize}. Page cache suffers a similar dilemma. I/O blocking writes to disk occur when a given percentage (e.g. 40~\%)
of total memory is occupied by dirty data. When the threshold is exceeded,
processes performing writes must wait for dirty data to be flushed to disk.

Furthermore, like memory, tmpfs and page cache are shared resources on a node.
If users on the node are heavily using memory to incite tmpfs to writes to swap space, or are performing data-intensive operations
that fill up the page cache, tmpfs/page cache performance on will be limited for other users. 
However, it is possible to request through the HPC scheduler a certain
amount of available memory. Ultimately, in-memory data will also need to be spilled to
disk if memory usage exceeds amount of available memory, although disk writes are 
likely to occur in tmpfs and page cache before requested available memory is filled.

\subsection{Scheduling Remarks}

A common recommendation in Spark is to limit the number of cores per 
executors to 5, to preserve a good I/O 
throughput (see \href{http://blog.cloudera.com/blog/2015/03/how-to-tune-your-apache-spark-jobs-part-2}{Cloudera blog}), 
but the rationale for this recommendation is hardly explained. We 
believe that throughput degradation observed with more than 5 cores per 
executor might be coming from full page caches.

Spark does not currently include any active management of disk I/Os or 
page caches. We believe that it would be beneficial to extend it toward 
this direction, to increase the performance of operations where local 
storage has to be used, such as disk spills or shuffles. For instance, 
workflow-aware cache eviction policies that maximizes page cache usage 
for the workflows could be investigated. 

An alternative Nipype plugin designed for running on Slurm was not used 
in the experiments. The Slurm plugin requests a Slurm allocation for 
each processed data chunk. Such a scheduling strategy was not ideal in 
our environment where oversubscription of nodes was not enabled.

Unlike Spark, Nipype by default opts to use all available CPUs rather 
than to load balance data across the cluster. That is, given 50 chunks 
and 40 cores, Spark will only use up 25 cores and process in two batches.
Nipype will also have no choice 
but to split up the processing into two batches, but will first process 
40 chunks, immediately followed by the remaining 10. While both are 
reasonable strategies for data processing, Spark may end up benefiting 
more from the page cache, as less data is written in parallel (25 vs 
40), giving more time for the page cache to flush.

Nipype's MapNodes, which apply a given function to each element in 
a list, were found to be slower than the Node, which apply a function 
to a single element, due to a blocking mechanism. For this reason, we 
selected to iterate through a series of Nodes in our code despite MapNodes being
easier to use.


\subsection{Other Comments}

Writing to node-local storage in a cluster environment comes at a cost, for both
Nipype and Spark without HDFS. When a node is lost, the node-local data is lost
with it. While Spark will recompute the lost partition automatically using lineage, 
Nipype will fail for all tasks requiring the data. Nevertheless, Spark will also fail if 
RDDs of node-local filenames are shuffled, as the data associated to the filenames
will not be shuffled with the RDD and there will be no mechanism in place to fetch it.

When executed on Lustre, Nipype will checkpoint itself, ensuring resumption from
last checkpoint during re-execution. This is particularly important in the case of compute-intensive 
applications, such as those found in neuroimaging. Spark also provides a checkpointing
mechanism, however, it requires HDFS.

It is common in neuroimaging applications for users to want access to all intermediate
data. Such a feature is currently only possible when writing to shared filesystem. 
It would also not be an option with Spark in-memory. To enable this, burst buffers or 
heterogeneous storage managers (e.g. Triple-H~\cite{islam2015triple}) could be used to ensure
fast processing and that all outputs (including intermediate outputs) will be 
sent asynchronously to the shared filesystem.

It was expected that Spark would experience longer processing times, particularly
with the small datasets, due to Java serialization. This was not found to be the case. 
Unlike Spark, Nipype performs a more thorough provenance capture, potentially owing to longer
processing times.

In this paper, we analyzed the effects of Big Data strategies on an map-only artificial
neuroimaging pipeline. This allowed us to examine the effects of these strategies
without being significantly obscured by other conditions. Studying the effects 
of such strategies on map-reduce type
workflows, in addition to real neuroimaging pipelines, would allows us to gain
further insight on the added
value of the Big Data performance stragies for neuroimaging use cases remains to be
done.

\section{Conclusion} 
\label{sec:conclusion}

Big Data performance optimization strategies help improve performance of typical 
neuroimaging applications. Our experiments indicate that overall, in-memory computing 
enables greater speedups than what can be obtained by using page cache 
and tmpfs. While page cache and tmpfs do give memory-like performance, 
they are likely to fill up faster than memory, leading to increased 
performance penalties when compared to in-memory computing. We conclude 
that extending Big Data processing engines to better support 
neuroimaging applications, including developing their provenance, 
fault-tolerance, and reproducibility features, is worthwhile.

Data locality plays an important role in application 
performance. Local disk was found to perform better than the shared 
filesystem despite having lower bandwidth, due to increased contention on 
the shared filesystem. Since local disk typically has less 
contention than shared filesystems, it is recommended to store data locally.
However, using local storage without a distributed file system 
may limit fault tolerance.

Although a more thorough analysis of lazy evaluation remains to be 
performed, it is speculated that this may be the cause of the general 
performance difference between Spark and Nipype. Furthermore, it was
found that lazy evaluation optimizations increase the likelihood of 
cache hits, thus improving overall performance.

Even though Big Data strategies are beneficial to the processing of 
large images, it is estimated that it would require running a 
functional MRI dataset with 280 concurrent subjects for any noticeable 
impact using our Lustre bandwidth estimate. Benchmarking Spark and 
Nipype using such a large fMRI dataset would be a relevant follow-up 
experiment, to test this hypothesis. It would also be useful to evaluate
other types of applications, such as ones containing data shuffling steps.

Finally, we plan to extend this study by including task scheduling 
strategies in a multi-tenant environment. We expect to observe 
important differences between Spark and Nipype, due to Spark's use of 
overlay scheduling. The impact of other Big Data technologies, such as 
distributed in-memory file systems(e.g.~\href{https://ignite.apache.org}{Apache Ignite}) and Lustre scalability issues~\cite{sparkhpc},
could also be investigated.

\section{Acknowledgments}

We are thankful to the Dell EMC HPC and AI Innovation Lab
in Roundrock, TX, for providing the infrastructure and high-quality
technical support.

\bibliographystyle{IEEEtran} 
\bibliography{biblio}

\end{document}